\documentclass[aps,prl,twocolumn,superscriptaddress]{revtex4-1}
\usepackage{amsmath}
\usepackage{color}
\usepackage{graphicx}
\usepackage{epstopdf}
\usepackage{color}
\begin{document}

\title{Interferometric measurement of arbitrary propagating vector beams that are  tightly focused}

\author{Pedro A. Quinto-Su}%
 \email{pedro.quinto@nucleares.unam.mx}
\affiliation{%
 Instituto de Ciencias Nucleares, Universidad Nacional Aut\'onoma de M\'exico, Apartado Postal 70-543, 04510, Cd. Mx., M\'exico.
}


\begin{abstract}
In this work we demonstrate a simple setup to generate and measure arbitrary vector beams that are tightly focused. The vector beams are created with a spatial light modulator and focused with a microscope objective with an effective numerical aperture of 1.2. 
The transverse polarization components ($E_x$, $E_y$) of the tightly focused vector beams are measured with 3 step interferometry. The axial component $E_z$ is reconstructed using the transverse fields with Gauss law. We measure beams with the following polarization states: circular, radial, azimuthal, spiral, flower and spider web. 
\end{abstract}
\maketitle

Structured beams and in general vector beams with arbitrary polarization states have been generated with different methods, including q-plates and programmable optical elements like spatial light modulators \cite{spiral, qplate, wangx, wang, liu, guo, mellado}. There are many potential applications of these beams, specially under tight focusing \cite{youngworth}. 
For example, beams with radial polarization have been used in $4\pi$ microscopy \cite{4pi1} to generate small spherical focused spots.

Measuring tightly focused fields has been a challenge due to the possibility of sub-wavelength structures in phase and amplitude in the 3 polarization components. 
Normally these beams are measured with nanoprobes \cite{R1d2} like in the case of SNOM \cite{microaxi}. 
Other studies have used optical microscopes to image the transverse polarization components (amplitude) of a reflected focused spot \cite{R1d7, exactmap}. 

Recently, we showed that it is possible to use classical interferometry to measure tightly focused propagating beams (amplitude and phase) that have linear or circular polarization before focusing \cite{interf1}. 

In this work, we demonstrate a general and simple method to measure tightly focused fields for any input polarization state, including arbitrary vector beams. The measurement requires 6 interferograms and two images of the transverse components to measure the tightly focused transverse fields, while the longitudinal component is reconstructed with Gauss law. 

\begin{figure}
\includegraphics[width=3.4in]{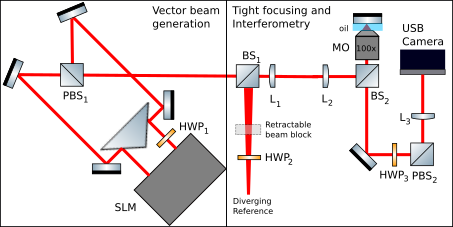}
\caption{Simplified experimental setup. Left: Vectorial beam generation. Right: Focusing, imaging and interferometry. 
Spatial light modulator (SLM), half wave plates (HWP), polarizing beam splitter cubes (PBS), beam splitter cubes (BS), lenses (L, $f_1=f_2=50\,$cm and $f_3=20\,$cm.) and microscope objective (MO).} 
\end{figure}
%
\begin{figure}
\includegraphics[width=3.3in]{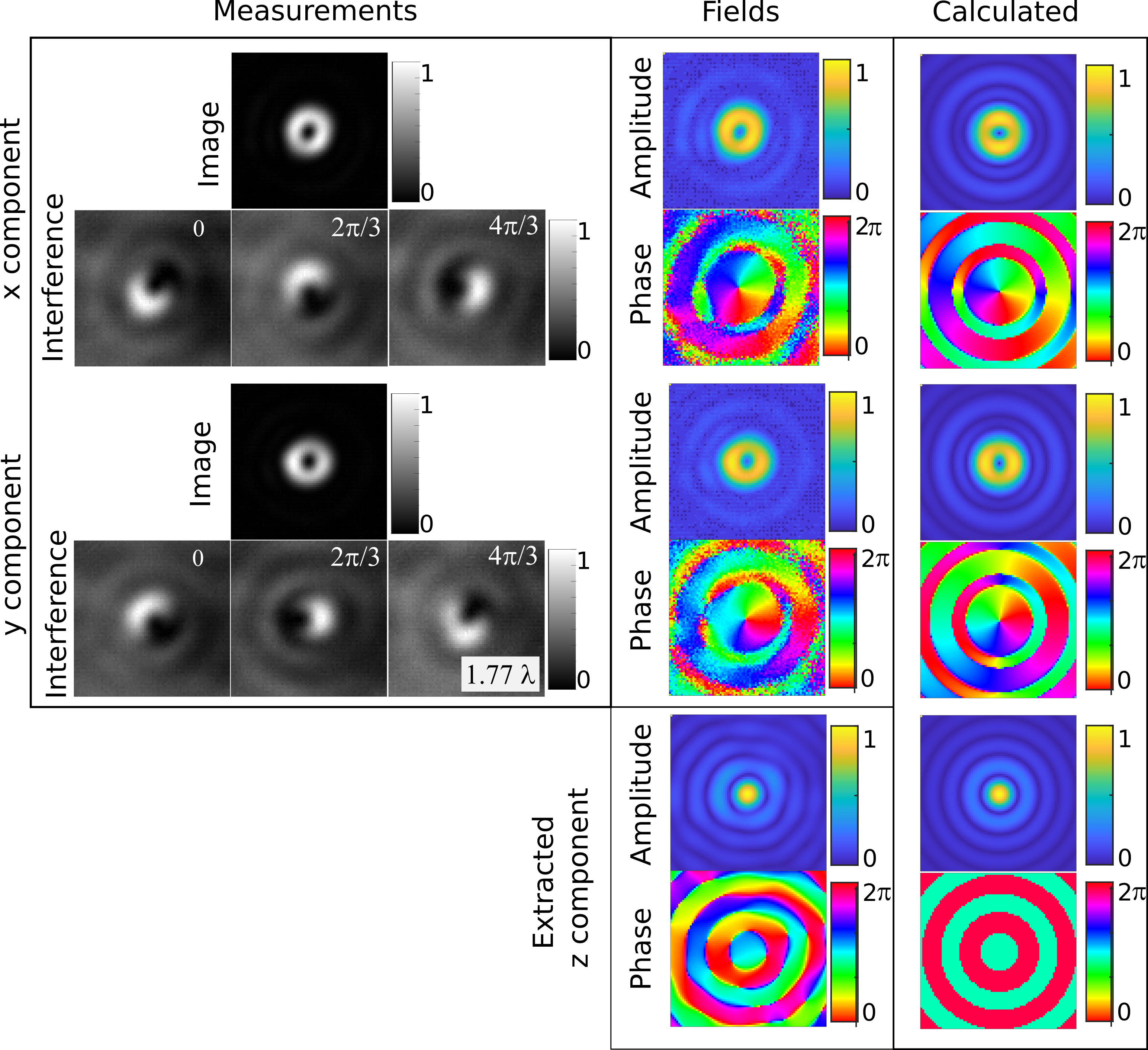} 
\caption{Interferometric measurements of the transverse field components and extraction of the axial polarization component for a vortex beam ($m=1$) with circular polarization (lhc). The left column has the measurements with the photograph of the beam on the top and the 3 interferograms at the bottom. The middle column has the amplitude which is the square root of the image and the phase is extracted from the interferograms (equations 1-2). The reconstructed axial field (equation 3). The last column has the calculated fields. Width of frames: $3.53 \lambda$.}
\end{figure}

\vspace{0.3cm}
{\it Experiment.} A schematic of the experimental setup is in Fig. 1. 
The laser has a wavelength of 1064 nm and it is divided into main and reference beams with a half wave HWP$_0$ plate and a polarizing beam splitter cube PBS$_0$ (not shown). 
The transmitted beam is the main beam and the reflected is the reference. 

The main beam is expanded in order to overfill the screen of a spatial light modulator (SLM) to generate a vector beam. This is shown in the left side of Fig. 1. The setup that produces the vector beam is an interferometer that combines both transverse polarization components and is similar to others that use spatial light modulators (SLM) \cite{wangx, wang, liu,guo}. The SLM modulates light with horizontal polarization, the screen is divided into two sections to independently modulate the components that make the vector beam. Light from the right side of the SLM propagates through a half wave plate (HWP$_1$) to rotate the polarization to a vertical state. Finally both components are combined with PBS$_1$. 

The reference beam propagates through an aperture and is focused with a lens $L_0$ (not shown), then it propagates through a HWP$_2$ (angle of 22.5$^o$) that rotates the polarization to a diagonal state with a phase difference of $\pi$ between the components. 
Both, main vector beam and diverging reference are combined with a beam splitter (BS$_1$) and then propagate through lens 1 (L$_1$). The distance from L$_1$ to the SLM is the focal length $f_1$. L$_1$ focuses the vector beam and collimates the reference beam. 
Then, the beams propagate through $L_2$ and are reflected with BS$_2$ into the back aperture of the microscope objective (MO, 100x, NA$=1.3$). 
Lens L$_2$ collimates the vector beam and projects the screen of the SLM into the back aperture of the MO. The reference beam is focused by L$_2$ and collimated with the MO. 

A mirror is positioned at the waist of the tightly focused vector beam with a piezo electric platform in order to image the beam \cite{R1d7, exactmap}. The imaging system is the MO, the projecting or tube lens (L$_3$) and the CMOS-USB camera. The effective magnification is the ratio of the focal lengths of $L_3$ and MO, resulting in a spatial calibration of $\Delta d=$47\,nm/pixel at the camera. 
The transverse polarization components (x,y) of the beam are selected with HWP$_3$ and PBS$_2$.

\vspace{0.3cm}
{\it Vector beam generation.} The digital holograms for both beams include a circular aperture that controls the effective numerical aperture of the system. We select an aperture radius of 2.2 mm that results in an effective NA of 1.2.
Both holograms modulate phase and amplitude \cite{boyd}. The amplitude of the beams shaped to a Gaussians with 
waists $w_0=6\,$mm that are larger than the aperture.

We generate vector beams with different polarization states: circular, radial, azimuthal, spiral, flower and spider web. 
The circular polarization states are generated by adding a phase (at the SLM) of $\pi/2$ to either the x component (rhc) or the y component (lhc) of the vector beam.
Radial polarization is obtained by modulating the amplitude of the x component with $\cos (\phi)$ and the y component with $\sin (\phi)$, where $\phi$ is the azimuthal angle at the SLM. Azimuthal polarization is produced modulating the x component with $\sin (\phi)$ and y with $\cos (\phi)$. Spiral polarization is a linear combination of the previous two states: x component modulated with $\cos (\phi)+\sin (\phi)$ and y with $-\cos (\phi)+\sin (\phi)$ \cite{spiral}. 

Flower and spiderweb \cite{flower} polarization patterns are created modulating x and y components with the following expressions: $\cos (s \phi /2)$ for x and $\sin (s \phi /2)$ for y, where $s$ is positive for flower and negative for spiderweb. Here we choose $s=\pm 8$.

\begin{figure*}
\includegraphics[width=6.8in]{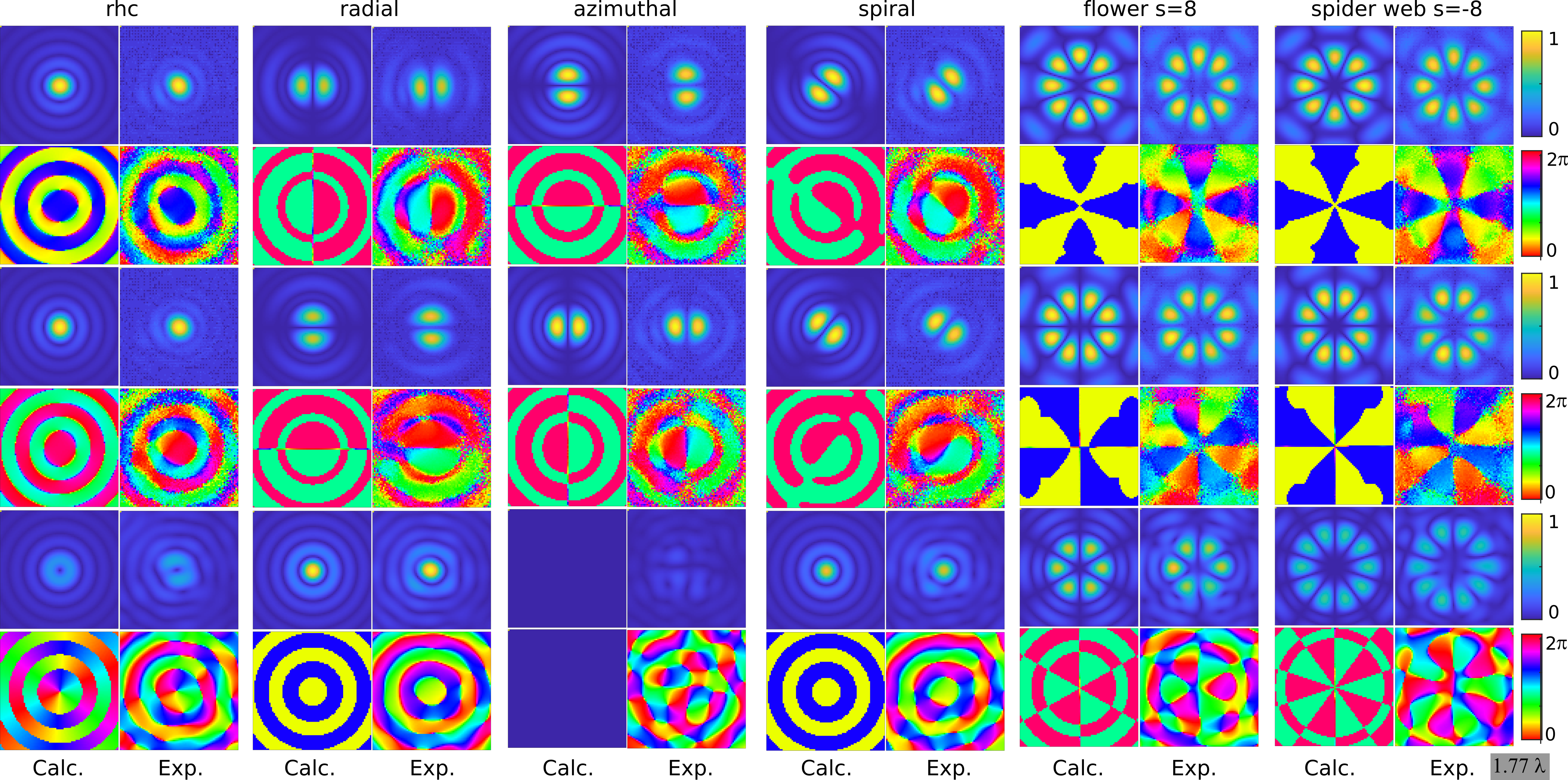} %
\caption{The results are arranged in column pairs for each vector beam, the left column is the simulated vector beam and the right has the measurements. The order (top to bottom) is: $|E_x|$, $\phi _x$, $|E_y|$, $\phi _y$, $|E_z|$, $\phi _z$. 
The tightly focused vector beams are circular polarization (rhc), radial, azimuthal, spiral, flower and spider web. The amplitudes are normalized with the global maximum between the 3 components. The scale bar is $1.77 \lambda$ and the width of the frame $3.53 \lambda$.} 
\end{figure*}

Before starting the measurements it is necessary to match the phases between the x and y components of the vector beam to the in-phase condition. This is done by measuring the phase between each transverse component and its corresponding reference. The phase difference between both components $\Delta \Phi = \Phi _x -\Phi _y$ is then added to the y component in order to compensate for small optical path length differences. 
The measurements are done with 3 step interferometry which is described next. 

\vspace{0.3cm}
{\it 3-step interferometry.} The transverse tightly focused fields are measured with 3-step interferometry \cite{zupancic}, where 3 interferograms $\mathcal{I}_{i,j}$ ($i=1-3$, $j=x,y$) are recorded between the main beam and the collimated reference for each transverse component. 
The $ith$ step phase shifts are added to the SLM: $\delta  \Phi _i =0$, $2 \pi /3$ and $4 \pi /3$ and the complex transverse fields ($E_x=A_x e^{i\Phi _x}$, $E_y=A_y e^{i\Phi _y}$) can be recovered with: 
\begin{equation}
E_x = -\frac{1}{3}(\mathcal{I}_{2,x} + \mathcal{I}_{3,x} - 2\mathcal{I}_{1,x}) + \frac{i}{\sqrt{3}} (\mathcal{I}_{2,x} -\mathcal{I}_{3,x})
\end{equation}
and
\begin{equation}
E_y = \frac{1}{3}(\mathcal{I}_{2,y} + \mathcal{I}_{3,y} - 2\mathcal{I}_{1,y}) - \frac{i}{\sqrt{3}} (\mathcal{I}_{2,y} -\mathcal{I}_{3,y}).
\end{equation}
A factor of $-1$ is added to the y component to compensate the $\pi$ phase difference between the components of the reference. We omitted the $(x,y)$ dependency to simplify the notation.

\vspace{0.3cm}
The measurement process is illustrated in Fig. 2 where the vector beam is a vortex of charge 1 with circular polarization (lhc). 3-step interferometry can recover both amplitude and phase for each transverse component. However, small power differences in the reference beam can result in incorrect power balance for the amplitudes. Hence, we first photograph each component (blocking the reference with a motorized beam block) of the beam which is proportional to the square of the amplitude $A_j ^2$ and the phase is extracted from the argument of equations (1-2). 

The left side of Fig. 2 contains the 8 measurements (4 per transverse component). The measurements are divided for each transverse component, with the image of the beam at the top (blocking the reference) and the 3 interferograms at the bottom. At the center column of Fig. 2 are the extracted amplitudes $A_j$ obtained from taking the square root of the beam image and the phase $\Phi _j$ is extracted from the interferograms.

\vspace{0.3cm}
Next, the z component ($E_z=A_z e^{i\Phi _z}$) is extracted from the x and y components \cite{interf1, progfocus} using Gauss law in the angular spectrum representation:
\begin{equation}
E_z=\mathbf{IFT}[-\frac{\mathcal{A}(k_x,k_y)}{k_z}(k_x  \mathbf{FT}[E_x]+\\
k_y \mathbf{FT}[E_y])]
\end{equation}
where $k_x$, $k_y$ are the angular coordinates at the SLM.

The aperture radius $R$ of $\mathcal{A}(k_x,k_y)$ is calculated  in k space \cite{fastfocal, progfocus}: $R=N \Delta d (NA)/\lambda $. 

The bottom part of the center column in Fig. 2 shows the z component extracted with eq. (3). The right column in Fig. 2 contains the calculated beam \cite{R1, fastfocal, novotnybook}. 

\begin{figure*}
\includegraphics[width=5.8in]{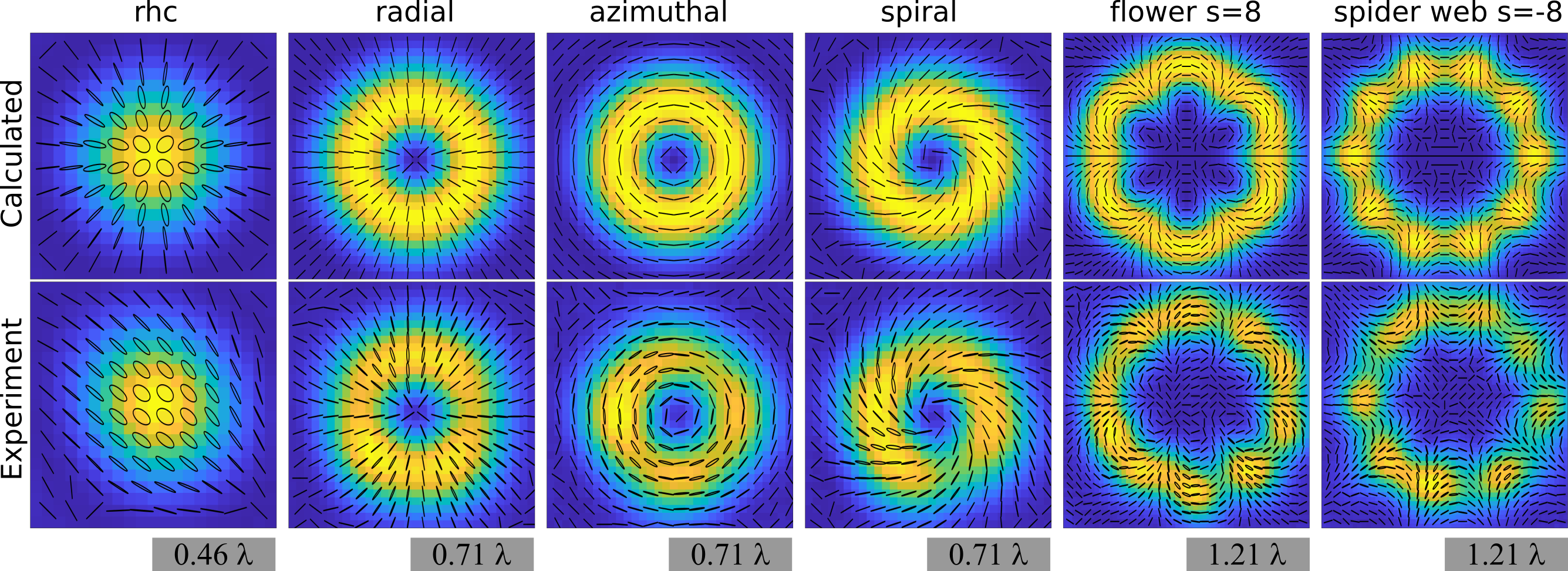} %
\caption{Polarization patterns for the different vector beams. The background is $S_0$.}
\end{figure*}

\vspace{0.3cm}
{\it Results.} Figure 3 contains the results for 6 different tightly focused vector beams. The results are arranged in column pairs, where the left column contains the calculated fields and the right the measured one. The quality is quantified with normalized cross correlations (NCC) \cite{ncc} which are essentially a dot product between the calculated field and the measured one. The definition is extended to the complex components so that $NCC=\frac{|E_{j, calc} E_{j, meas}^*|}{\sqrt{|E_{j, calc}|^2} \sqrt{|E_{j, meas}|^2}}$.

The first case is a beam with circular polarization (rhc), where we observe the expected spin-orbit coupling in the z component in the form of an optical vortex with charge equal to one. The NCCs are $(0.87,0.89,0.84)$ for the x,y and z components respectively. 
The beam with radial polarization has a large z component with NCCs of $(0.91,0.89,0.89)$. The azimuthal polarization case should only have x and y components with a vanishing z component. However, in the experiment we get a residual z component with $2\%$ of the total power. 
The NCCs are $(0.9, 0.9)$ for x and y components.
The spiral polarization case has NCCs of $(0.9,0.87,0.84)$. Finally the cases of flower and spiral polarizations with $s=\pm 8$ are in the last two column pairs. The NCCs are $(0.88,0.84,0.83)$ and $(0.9,0.86,0.82)$ for flower and spider web respectively. 

\vspace{0.3cm}
{\it Stokes parameters.} The Stokes parameters are extracted directly from the measured transverse fields: $S_0=|E_x|^2+|E_y|^2$, $S_1=|E_x|^2-|E_y|^2$, $S_2=2 Re (E_x E_y^*)$, $S_3=-2 Im (E_x E_y^*)$. 
The transverse polarization patterns \cite{galvez} with $S_0$ are in Fig. 4. The most complex patterns are the flower and spider web and the experimental patterns are close to what is expected, in other cases there is a slight ellipticity in some areas like in the case of the spiral beam. 

\vspace{0.3cm}
{\it Conclusion.}
We obtain reasonable agreement between the measured fields and the calculated ones, with the smallest NCC of 0.82.  This work shows that it is possible to measure arbitrary vector fields in a very simple way with classical interferometry. 
There are many possibilities for the interferometer, including in-line configurations and other modifications that might be convenient for different experimental setups.

\subsection*{Acknowledgements}
Work partially funded by DGAPA UNAM PAPIIT grant IN107222, CTIC-LANMAC and CONACYT LN-299057. 
Thanks to Cristian Mojica Casique for technical support and Jos\'e Rangel Guti\'errez for fabricating optomechanical components. 

\bibliographystyle{apsrev4-1}


\end{document}